\documentclass[conference]{ieeeconf}

\usepackage{cite}
\usepackage[pdftex]{color}
\usepackage[pdftex]{graphicx}
\usepackage{alltt}
\usepackage{url}
\usepackage{xspace}
\usepackage{here}
\usepackage{listings}
\usepackage{multirow}

\usepackage{my}

\begin{document}

\title{Designing a CPU model: from a pseudo-formal document to fast code}

\author{%
\parbox{15cm}{\centering
  F. Blanqui$^{1,2}$, C. Helmstetter$^{1,2,\dag}$, V. Joloboff$^{1,2}$, J.-F. Monin$^{1,3,4}$ and X. Shi$^{1,4}$\\[1ex]
  {\small\itshape $^1$LIAMA-FORMES, $^2$INRIA, $^3$CNRS, $^4$Universit\'{e} de Grenoble 1}\\
  $^\dag$claude.helmstetter@inria.fr
}
}
% \institute{INRIA-LIAMA}
% \institute{INRIA-LIAMA
% \and Universit\'{e} de Grenoble 1
% \and CNRS-LIAMA}
% \author{First Author, Second Author and Third Author% <-this % stops a space
% 	\\\textit{Author Affiliation}%
% 	\\\textit{Author Email Addresses}
% }
% For authors with multiple affiliations, use multiple centered parboxes, as follows...
%\author{ \parbox{2 in}{\centering Huibert Kwakernaak\\
%         \textit{University of Twente}\\
%         \textit{h.kwakernaak@autsubmit.com}}
%         \hspace*{0.5 in}
%         \parbox{2 in}{ \centering Pradeep Misra\\
%         \textit{Wright State University}\\
%         \textit{pmisra@cs.wright.edu}}
%}

\maketitle

\begin{abstract}
  For validating low level embedded software, engineers use simulators
  that take the real binary as input. Like the real hardware, these
  full-system simulators are organized as a set of components. The main
  component is the CPU simulator (ISS), because it is the usual
  bottleneck for the simulation speed, and its development is a long
  and repetitive task. Previous work showed that an ISS can
  be generated from an Architecture Description Language (ADL).  In
  the work reported in this paper, we generate a CPU simulator
  directly from the pseudo-formal descriptions of the reference
  manual. For each instruction, we extract the information describing
  its behavior, its binary encoding, and its assembly syntax. Next,
  after automatically applying many optimizations on the extracted
  information, we generate a SystemC/TLM ISS. We also
  generate tests for the decoder and a formal specification in
  Coq. Experiments show that the generated ISS is as fast and stable
  as our previous hand-written ISS.
\end{abstract}

%%%%%%%%%%%%%%%%%%%%%%%%%%%%%%%%%%%%%%%%%%%%%%%%%%%%%%%%%%%%%%%%%%%%%%%%%%%%%
\section{Introduction}

Developing a new System-on-Chip (SoC) for some embedded systems requires the
design of {\em abstract models}~\cite{tlm-book}. These models ease the design
and the validation by providing a global view of the future system, allowing to
certify protocols, simulate the embedded software, and decide the correctness of
hardware executions.

Like the real hardware, a model of a full system is organized as a set of
components. When a system is simulated, most of the computation time is spent in
the component modeling the processor. This component is called an {\em ISS}
(Instruction Set Simulator). Fast simulations require to implement many
optimizations techniques in the ISS, such as {\em dynamic
  translation}~\cite{dyntrans}.  Even without optimizations, writing an ISS is a
long, tedious then error-prone task because functional specifications of processors
are generally over 500~pages long. % JF: unsure of grammar

Reference manuals of processors are mainly written in natural language, but some
parts are described with pseudo-formal descriptions that can be automatically
parsed and interpreted. In this work, we present how to take advantage of these
pseudo-formal sections in order to generate automatically most of the code of a
CPU model.

In addition of automatic extraction of the pseudo-formal sections, we take the
most of the intermediate representation, which can be handled easily by
software, to apply many kinds of analysis and optimizations. Our goal is to
generate an ISS that is as good as an hand-written one, without any manual
modification of the generated code.

We consider the ARMv6 architecture, which is implemented by the ARM11 processor
family. The reference manual~\cite{arm6refman} of the CPU part (i.e., excluding
the memory management part) counts 617~pages. This manual is mainly written in
natural language, but each instruction is described by three elements that can
be automatically parsed:
\begin{itemize}
\item a table describing the instruction binary encoding
\item a piece of {\em pseudo-code} describing its behavior
\item the syntax of the instruction in assembly code.
\end{itemize}

We have developed a tool chain that extracts the pseudo-formal parts of the
ARMv6 manual, to an easy-to-use intermediate representation. In this
intermediate representation, the instruction behavior is represented by {\em
  abstract syntax trees} (ASTs). Next, we developed a set of back-ends. In
addition to a formal specification and unit tests, we generate a fast C/C++ ISS,
which is part of the SimSoC open-source project since the 0.7 release.

Note that the PowerPC, MIPS, and SH2 architecture reference manuals use a
similar structure to describe the instructions (i.e., encoding plus syntax plus
pseudo-code). The PowerPC and MIPS may be a little more complicated to parse
because they use non-ASCII characters inside the pseudo-code. On the contrary,
parsing SH2 documentation should be easier because it uses a simple subset of C
to describe the instruction behavior.

Our generated ISS uses the same optimization techniques as classic
hand-written ISSes, such as dynamic translation. Dynamic translation means that
the result of the decoder is stored, to avoid the decoding time if the same
instruction is executed again.
The dynamic translation technique we use is described in~\cite{ossc09}.
% An instruction can be stored as a C data structure, or
% compiled to native code.  The compilation to native code is done using the LLVM
% runtime compiler~\cite{llvm}, and only for blocks which are executed very often.
Generating such an optimized ISS requires to translate the pseudo-code to C/C++, but
also to collect new data, such as the lists of parameters and local variables,
and to combine data from different sources (see for example the {\stt
  may\_branch} function discussed in section~\ref{s:maybranch}).

% TODO: it is not just a translation. collect and combine data.

This article is structured as follows. Section~\ref{s:rw} is devoted to related
work. Extraction of interesting parts from the manual, code transformations
needed for correctness and better performance, and code generation are described
in section~\ref{s:gen}. Section~\ref{s:result} presents the results.

%%%%%%%%%%%%%%%%%%%%%%%%%%%%%%%%%%%%%%%%%%%%%%%%%%%%%%%%%%%%%%%%%%%%%%%%%%%%%
\section{Related Work} % => Vania
\label{s:rw}

Previous work proposed solutions to generate an ISS from an Architecture
Description Language (ADL), mainly for retargetability issues. In the
JACOB system~\cite{Leupers99} a processor is described with the MIMOLA
language, a low level description. From the MIMOLA input, a C program
is generated that simulates the processor. The MIMOLA compiler
generates C macros using the basic functions have been manually coded
however.
% So it is not entirely generated, there are basic operations
% manually coded.

In~\cite{Engel:2000:GTS:334012.334036} the processor is described using
a {\em Processor Description File}, but this is more a kind of
pre-processor, as the language contains C macros like constructs.  This
work is doing static compiling simulation, it does not generate a
simulator, it generates a native program for the host computer
simulating the binary on the target processor.

The FACILE language~\cite{facile-schnarr-2001} can be used to generate
simulators.
% But it is not a declarative language, it includes specific
% procedural simulation constructs. The simulator is an infinite loop
% calling the individual functions.
The generated simulator has optimization using partial evaluation techniques
similar to the specialization optimization described in the next
section. However, they target a low abstraction level, more suitable to
performance evaluation than functional validation.
Similarly, using the MADL language, the authors of~\cite{Qin:2004:FCM:998300.997171}
have generated a cycle accurate simulator for several architectures.

Other researchers have used a kind of virtual machine approach, where the
processor instruction set is described in terms of the basic operations. The
LISA language \cite{naul-braun-jit-iccs} uses this approach. Like FACILE, this
work targets a lower abstraction level than us.
% From the processor description, it generates code but this
% code has to be linked with manually written programs.

The virtual machine approach is also used in the QEMU simulator~\cite{QEMU}
which has been manually coded.

An approach closer to our work is the EXPRESSION-ADL language~\cite{1151083}.
It generates a decoder and static compiled simulator for the target
instructions. They report a maximum speed of 15 Mips for ARMv4, whereas we
simulates the ARMv6 architecture at 90 Mips or more on several benchmarks.

% They generate C++ templates and the
% late template instantiation makes it possible to optimize the
% generated code. This is similar to the flattening and sub-expression
% optimizations that we achieve.

In the works mentioned above, some of the systems do generate decoders
for the target binary, and some do not.  In the work presented here,
we generate the simulator, the decoder and some additional tests.

% Moreover, for all the above mentioned work, quoted performance of the
% dynamic compiled or interpreted simulators is ranging from thousands
% of instructions per second to at best about 15 Millions instructions
% per second for an ARMv4 architecture. Our generated simulator simulates
% an ARMv6 architecture and runs at 90 Mips or more on several
% benchmarks.

Zhu and Gajski \cite{staticISS-zhu-gajski} have done a static
compilation retargetable simulator. The input instructions are first
translated into a virtual machine instructions, with an infinite
number of registers. Then a back end translates the virtual
instructions into host code, using a dedicated register allocator.
The result is a compiled program running on the host.  Because it is
static compilation, it is fast, with benchmarks running at 200 Mips
(on a 2004 PC).
We use dynamic translation instead of static compilation, because
static compilation is not suitable for dynamically loaded code.
\section{The ISS Generator}
\label{s:gen}

\begin{figure*}\centering
\scalebox{.5}{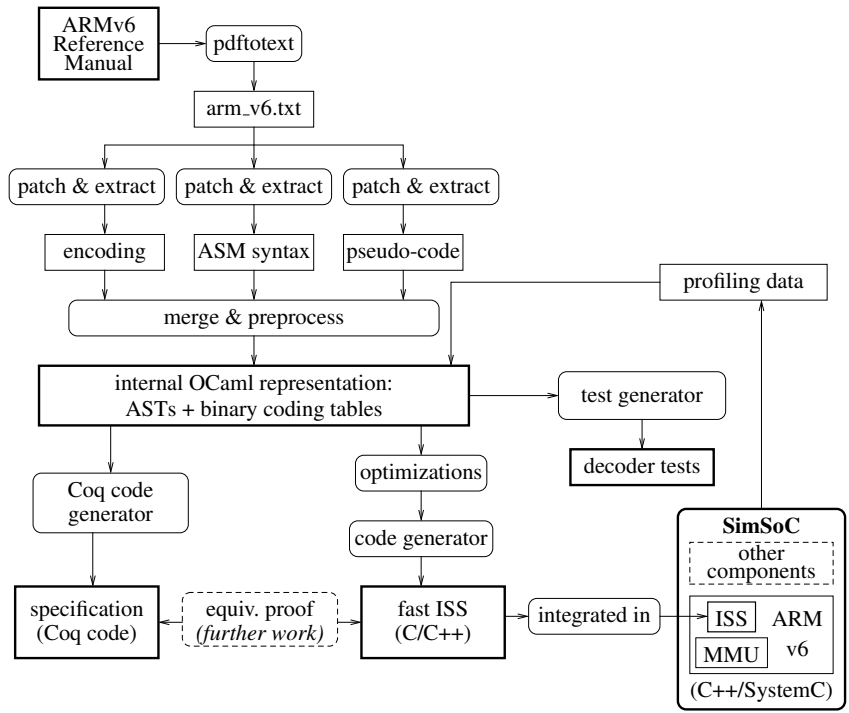 }
\caption{Overall Architecture}
\label{fig:archi}
\end{figure*}

Starting from the ARMv6 architecture reference manual (reference: {\stt ARM DDI
  0100I}), we have built a tool chain composed of a front-end that extracts and
parses the pseudo-formal parts, and of several back-ends. The main back-end
generates a C/C++ ISS suitable for fast simulations. Another back-end generates
tests and a last one generates a formal specification in Coq, which can be used
to develop proofs.  Fig.~\ref{fig:archi} describes the overall architecture.
% JF: was unsure that "conduct proofs" is acceptable

\subsection{Extraction and parsing}

% pdftotext, patch, and extraction
The very first step is to run the command {\stt pdftotext}. The
resulting file is 28500 lines long (excluding parts B, C, and D,
which are not related to the CPU). Next, we extract three smaller files,
each containing one kind of information: a 2100 lines-long file
contains the pseudo-code, another 800 lines-long file contains the
binary encoding tables, and the ASM syntax file is 500 lines
long. Each extraction is done by a small ad-hoc OCaml program. Before
extraction, a patch is applied to the main text file (using the Unix
{\stt patch} command). This patch fixes some obvious mistakes coming
from the original document, such as misspelling in a function name,
unclosed parenthesis (Thumb STMIA instruction), missing line
(condition check of the {\stt CLZ} instruction), etc.

% parsing
Next, each extracted file is parsed with the corresponding parser. The most
complicated parser is for the pseudo-code. Two preliminary phases solve
issues related to line breaks and indentation, given that indentation
defines the blocks in a Python-like way. Then, a classical lexer-parser
combination builds the {\em abstract syntax trees} (ASTs).  The whole extraction
and parsing task is performed by 1400 lines of code (OCaml, OCamlLex, and OCamlYacc).

% manual extraction of validity constraints
As the reader can notice, we extract only 10\% of the document. Part of the
information that is not extracted is useless, such as the typical use cases
for instructions. There are redundancies also. However,
important pieces of information must still be taken into account.
In particular, many instructions have additional {\em
  validity constraints}, which are described only in informal English text. For
example, the {\stt Rn} register of the {\stt UXTAH} instruction must not be {\stt
  R15}. Because this information is required by code generators, we have
extracted all validity constraints by hand to an OCaml file (300~lines of data).

% other data
Finally, there is some important information that is neither extracted nor
needed by the generator; the corresponding C/C++ code is written by
hand. Actually, we use a generator only for parts that are related to the
instruction list (255 entities described in a same way). Indeed, it is not worth
using a generator for something that is not repetitive, because the generator
would be longer than the generated code.

\subsection{Transformations and optimizations}

Before generating the code for the fast SystemC/TLM ISS, several transformations
and analysis are applied to the OCaml internal representation. There are two
categories of transformations: some are required for the correctness of the
generated code, others improve the simulation speed without modifying the behavior.
We present some of these transformations in the remaining of this sub-section:
three of each category. All the transformations are implemented by a total of
1200 lines of code.

\subsubsection{Symbolic expression as parameter}

The reference manual contains some pseudo-code that looks like this:

{\small
\begin{alltt}
V Flag = OverflowFrom(Rn + shift_op + C Flag)
\end{alltt}
}

This is not a function call taking one integer as a parameter, but a function that
takes the symbolic expression {\stt Rn + shift\_op + C Flag} as a
parameter. Indeed, computing the overflow bit requires to know which operator is
used (addition or subtraction), and each operand value.  We recognize such calls
in the ASTs and replace them by:

{\small
\begin{alltt}
V Flag = OverflowFrom\underline{Add3}(Rn\underline{,}shift_op\underline{,}C Flag)
\end{alltt}
}

Similar transformations are applied to {\stt CarryFrom}, {\stt BorrowFrom}, and
{\stt SignedSat}.

\subsubsection{Addressing mode variants}

Many ARM instructions are described in two parts: the instruction body and the
{\em addressing mode} or {\em shifter operand}. For example, the third argument
of an addition can be either an immediate value, a register, or a shifted value.
Each addressing mode case is described in the same way as a normal
instruction, with an encoding table, a syntax, and a piece of pseudo-code.

One difficulty is that some instructions, such as {\stt SRS} and {\stt RFE}, use
a variant of the addressing mode pseudo-code. For example, after extraction and
parsing, we know that the pseudo-code of {\stt IA} addressing mode should be:

{\small
\begin{alltt}
start_address = Rn
end_address = Rn+(NbOfSetBitsIn(reglist)*4)-4
if ConditionPassed(cond) and W==1 then
    Rn = Rn+(NbOfSetBitsIn(reglist)*4)
\end{alltt}
}

However, the textual description of {\stt SRS} explains that there are some
differences in the above code when this code is used with this particular
instruction:

\begin{itemize}
\item {\stt Rn} is replaced by the banked version of register {\stt R13} for a
  mode given as parameter.
\item the register list length is 2.
\end{itemize}

Thus, we provide a remedial patch function that applies these two transformations.
% when necessary.
 Another patch function fixes the {\stt RFE} instruction. Note that
these patch functions are very simple. Indeed, the {\tt SRS} patch function looks like:

{\small
\begin{alltt}
let code1 = replace_exp code0
{\rm\em (* replace... *)}(Reg (Var "n", None))
\hspace*{1mm} {\rm\em (* ... by... *)} (Reg (Num "13", Some (Var "mode")))
in let code2 = replace_exp code1
        (Fun "NbOfSetBitsIn", [Var "reglist"])
        (Num "2") in...
\end{alltt}
}

\subsubsection{Register write-back and data aborts}
\label{s:writeback}

Load and store instructions read the base address from a register. Under some
conditions, the contents of this base register is incremented or decremented by
an offset. In the extracted pseudo-code, this \emph{write-back} to the base
register is done before the memory access itself. However, if the memory access
fails and thus raises a {\em data abort} exception, then the base register must
keep its original value. As for addressing mode variants, this rule is only
explained using informal text.

Our generated ISS manages {\em data aborts} using C++ exception mechanism. As a
consequence, moving the statement doing the write-back at the end of the
instruction code (and so after any possible {\stt throw}) is sufficient to keep
the base register unchanged in case of exception. Considering the IA pseudo-code
shown above, we need to move this statement:

{\small
\begin{alltt}
if ConditionPassed(cond) and W==1 then
    Rn = Rn+(NbOfSetBitsIn(reglist)*4)
\end{alltt}
}

It is not as simple as it looks, because some instructions such as {\stt LDM(3)}
modify the processor mode, thus changing the meaning of ``{\stt Rn}''. In this
case, the write-back must affect a banked version of {\stt Rn} instead of the
current version.

\subsubsection{Instruction flattening}

The initial intermediate representation contains elements that describe either
an instruction or an addressing mode case. For each instruction $A$ that can
use an addressing mode $B$, we generate a new instruction $AB$, where the data
structures of $A$ have been instantiated with the data of $B$. The benefits are
twofold: the following generation steps become simpler; and the generated code
is faster. After this transformation, called {\em flattening}, we have slightly
more than twice as many instructions.

The flattening step operates on the four elements:
\begin{itemize}
\item The mode case pseudo-code is inlined at the beginning of the instruction
  pseudo-code.
\item The validity constraint lists, which have been extracted manually, are
  appended.
\item The ASM syntax of the instruction contains a special parameter that must
  be replaced by the mode case syntax. For example, the syntax: \\
  {\stt ADC\{$\langle$cond$\rangle$\}\{S\}
    $\langle$Rd$\rangle$,$\langle$Rn$\rangle$,$\langle$shifter\_operand$\rangle$},\\
  combined with the syntax: {\stt $\langle$Rm$\rangle$,LSL\#$\langle$shift\_imm$\rangle$},\\
  yields the flattened syntax:\\
  {\stt ADC\{$\langle$cond$\rangle$\}\{S\}
    $\langle$Rd$\rangle$,$\langle$Rn$\rangle$,$\langle$Rm$\rangle$,LSL\#$\langle$shift\_imm$\rangle$}.
\item The encoding tables are merged, keeping the most specific option for each
  bit: a constant replaces a parameter, short parameters replace long
  parameters. An example is given by Fig.~\ref{fig:flatten}.
\end{itemize}

\begin{figure*}\centering
\begin{tabular}{|c|c|c|c|c|c|c|c|c|c|}
\multicolumn{10}{c}{\small\em (a) binary encoding of the {\stt ADC} instruction}\\
\hline
31 $\ldots$ 28 & 27 26 & 25 & 24 \dotfill 21 & 20 & 19 $\ldots$ 16 & 15 $\ldots$ 12 & \multicolumn{3}{c|}{11 \dotfill 0} \\\hline
\stt cond & \stt 0~0 & \stt I & \stt 0~1~0~1 & \stt S & \stt Rn & \stt Rd & \multicolumn{3}{c|}{\stt shifter\_operand} \\
\hline
% \multicolumn{10}{c}{~}\\
\multicolumn{10}{c}{\small\em \phantom{\LARGE I}(b) binary encoding of the ``logical shift left by immediate'' operand\phantom{\LARGE I}}\\
\hline
31 $\ldots$ 28 & 27 26 & 25 & 24 \dotfill 21 & 20 & 19 $\ldots$ 16 & 15 $\ldots$ 12 & 11 \dotfill 7 & 6 $\ldots$ 4 & 3 $\ldots$ 0 \\\hline
\stt cond & \stt 0~0 & \stt 0 & \stt opcode & \stt S & \stt Rn & \stt Rd & \stt shift\_imm & \stt 0~0~0 & \stt Rm \\
\hline
% \multicolumn{10}{c}{~}\\
\multicolumn{10}{c}{\small\em \phantom{\LARGE I}(a+b) resulting binary encoding of the flattened instruction\phantom{\LARGE I}}\\
\hline
31 $\ldots$ 28 & 27 26 & 25 & 24 \dotfill 21 & 20 & 19 $\ldots$ 16 & 15 $\ldots$ 12 & 11 \dotfill 7 & 6 $\ldots$ 4 & 3 $\ldots$ 0 \\\hline
\stt cond & \stt 0~0 & \stt 0 & \stt 0~1~0~1 & \stt S & \stt Rn & \stt Rd & \stt shift\_imm & \stt 0~0~0 & \stt Rm \\
\hline
\end{tabular}

\caption{Flattening the ADC instruction with the shift left by immediate operand}
\label{fig:flatten}
\end{figure*}

\subsubsection{Pre-computation of static sub-expressions}

Because we use dynamic translation, an instruction is generally decoded once and
executed many times. So, if a sub-expression depends only upon the
instruction parameters but not the processor state, we can accelerate the simulation
by moving this sub-expression from the execute function to the decoder.

For example, the {\stt IA} addressing mode contains the sub-expression {\stt
  NbOfSetBitsIn(reglist)*4}. The variable {\stt reglist} is a 16-bits value
stored in the instruction encoding, and {\stt NbOfSetBitsIn} is a pure function
(Hamming weight), so it can be pre-computed. An optimization function replaces
this expression by a new parameter {\stt nb\_reg\_x4}, and additional code is
generated in the decoder to compute it.

We provide manually a list of patterns to be pre-computed to the OCaml
optimizer, which then applies automatically the transformations.

\subsubsection{Specialization}

The ARM instructions accepts generally many options and flags. The
{\stt ADC instruction} takes on optional condition and the flag {\stt
  S} decides whether the status register must be updated. Checking
whether the options are present takes a little time at each execution,
whereas most of the time the options are absent (i.e., the condition
is {\em ``Always''} and {\stt S} is false).

Instead of generating one generic {\stt ADC} instruction, we generate many
specialized instructions. Firstly, we duplicate the pseudo-code: in one version,
we replace the specialized flag (e.g., {\stt S}) by {\stt 1}, and in the other
we replace it by {\stt 0}. A simple subsequent pass removes the obvious dead
code, knowing that more complicated optimizations will be done by the C++
compiler itself. Furthermore, for each conditional instruction, we generate an
unconditional variant, in which the condition check is removed.

Specialization can increase dramatically the number of instructions, and thus
the size of the generated code. Among the consequences, the compilation time may
become huge and the ISS binary load time will increase. To solve this problem,
we simulate some benchmarks on the generated ISS, record how many time each
instruction is executed, and inject these data back into the ISS
generator. Thus, the specialization pass knows each instruction weight and
whether it is worth to specialize it.

\subsection{Code generation}

The following elements are generated and included in the final ISS:
\begin{enumerate}
\item The types used to store an instruction after decoding.
\item Two decoders: one for the main ARM instruction set and another for the
  Thumb instruction set.
\item The semantics function, corresponding to the extracted and optimized
  pseudo-code. This is quite straightforward because pseudo-code ambiguities
  have been solved before.
\item The {\stt may\_branch} function that detects basic block terminators
  (i.e., branch instructions).
\item The ASM printers, used to print debug traces.
% \item (Part of) an ARM to LLVM translator.
\end{enumerate}

An instruction is stored in a {\stt struct} type containing an identifier, a
pointer to the semantics function, and an {\stt union} field containing the
instruction parameters. The list of the instruction parameters is computed
automatically by analysing the pseudo-code. There are 80 distinct parameter
lists; for each of them, we generate a {\stt struct} type, which is referenced
in the {\stt union} field.

The generation of the decoder is somewhat tricky, due to some features of the
ARM instruction set. In particular, looking at the encoding tables is not
enough, because some binary words match many instruction tables. For example, any
{\stt UXTH} instruction matches the encoding tables of {\stt LDRB}, {\stt
  LDRBT}, and {\stt UXTAH}; trying to decode the addressing mode eliminates the
{\stt LDRB} and {\stt LDRBT} candidates, and checking the validity constraint of
{\stt UXTAH} ({\stt Rn$\ne$R15}) eliminates this third candidate.  To solve this
issue, our decoders work in two phases: the first phase selects the candidates
using a {\stt switch} statement, the second phase evaluates the validity
constraints.

\label{s:maybranch}
The dynamic translation technique we use requires to recognize the {\em basic
  blocks} (i.e., a sequence of instructions always executed in a row). An
instruction is a basic-block terminator if it {\em may branch} for some states
of the processor. Some particular instructions are managed manually, but the
{\em may-branch} condition is computed automatically for most instructions, using a
small static analyser.

We illustrate the principle of this static analyser on the {\stt LDR}
instruction combined with the {\em register pre-indexed} addressing mode. First,
we go trough the AST, searching for assignments to the PC, which, on ARM, is the
register {\stt R15}. We encounter three register assignments: 1) {\stt
  Rn=address} (code of the addressing mode), 2) {\stt PC=data AND...}, 3) {\stt
  Rd=data}. Then, we associate a condition to each assignment. From the first
assignment, we conclude that the instruction may branch if {\stt n==15}. The
second (respectively third) assignment appears on the {\stt then} side (resp.
{\stt else} side) of an {\stt if} statement whose condition is {\stt Rd is R15}.
So the assignment 2) yields the condition {\stt d==15}, and the latter yields
the condition {\stt false} (obtained by reduction of {\stt d==15\&\&d!=15}). At
this point we have the global {\em may-branch} condition {\stt
  n==15||d==15}. However, one of the validity constraint associated with this
instruction is {\stt n!=15}, so the final {\em may-branch} condition is
simplified to {\stt d==15}. This condition will be evaluated at decode-time once
the value of {\stt d} in known, deciding whether it is the end of the current
basic-block.

% TODO: add a detailed example (LDR), showing that we use data from VC list

% C instead of C++ for formal verification

% LoC: generator 1200, generated: TODO
The whole OCaml extractor-generator is 5400
lines long, and it generates 74,000 lines of C/C++. Note that the ISS
is mainly written in C because we have started some formal
verification work using tools not compatible with C++.

To allow full system simulation, we integrated our ISS in a SystemC/TLM module
(4000~LoC in addition of the generated code), and added this module to the
open-source SimSoC project~\cite{ossc09}. The generated code is already released
as part of the version 0.7 of
SimSoC\footnote{SimSoC URL:\url{http://gforge.inria.fr/projects/simsoc/}}, and
we plan to release the generator itself in a near future.

% stand-alone minimal simulator

Additionally, we generate a formal specification of the ARMv6 in Coq, which is
used by another work investigating the certification of simulators. The Coq code
generation uses the same techniques than the C++ generator, excepted that
optimizations are disabled, and imperative code is translated to functional code
using monadic specifications~\cite{wadjfp09}. Our Coq specification is similar
to the HOL specification of~\cite{FoxM10}.
% ARMv7 architecture presented in~\cite{FoxM10}
% is manually generated, is contrast with ours.

\subsection{Automatic test generation} % => Shi Xiaomu

In order to validate the SimSoC decoder, we prepare massive binary tests.
We built an automatic test generator that generates all possible instructions
which are neither undefined nor unpredictable.
%With the current existing tests, we can not say that the decoder can decode every
%instructions correctly, but the one appears in the tests.
%Using this generator, we are able to cover all predictable cases.
%It will be more convenience for validating our decoder.
%In the meantime,
We generate two files. The first contains the instruction binary, in the ELF
format. The second contains the expected assembly code.
%   Another file which
% contains the same tests in assembly code is generated as well.
% It is going to be compared with the result of decoding the binary tests.
% The binary instructions are generated according to the instruction encoding
% as specified in the reference manual.
% The syntax of the ARMv6 assembly language is specified for each instruction.
Both files are generated according to the instruction encoding and syntax
as extracted from the reference manual.

% It is also patched and extracted from the reference, we have mentioned in section ~ref{}.
% The first step is to choose a value for each instruction parameter.
% variables occurring in an instruction, according to the reference manual.
%we know which variables have an impact on the generated binary and assembly codes.
% For example, the parameters of the {\stt ADC} instruction (see
% Fig.~\ref{fig:flatten}) are {\stt Rd}, {\stt Rn} and {\stt shift\_imm}.
% the flattened ADC instruction encoding,
% Instruction are producted with different combinations of values for them.
% The table stores all possible values for them. % of the instructions.
The parameter values are chosen with respect to the validity constraints to
ensure that the instruction is defined and predictable.
% Some validity
% constraints are dealt with during the parameter generation.
The validity
constraints are dealt with during the parameter generation.
For example, {\stt
  Rn} in instruction {\stt LDRBT} cannot be {\stt R15}, so we chose directly a value
between 0 and 14.
% Some other validity constraints, which involve two or more
% parameters at the same time, are handled in a second stage.
% If there is no this kind of constraints assigned to
% a instruction, a value will directly selected from the value table.
% Otherwise, it will select value according to the constraint.
Continuing the example of {\stt LDRBT}, another constraint states
that {\stt Rd} and {\stt Rn} must be different:
the generator produces two different values from the previous
table and assigns them to {\stt Rd} and {\stt Rn}.
% The remaining selected values are used to form the binary code
% and the same assembly code together.

The generated binary instructions are given as input to the SimSoC
decoder. The latter prints the corresponding assembly code which is then
compared with the generated assembly code using the Unix command {\stt
  diff}.  Minor issues have been detected and fixed in this way. We did also
compare the result with the one of the GNU disassembler ({\stt arm-elf-objdump
  -d}), but as the GNU syntax is slightly different, comparison must be done by
hand.

\section{Reliability and Performances}
\label{s:result}

\subsection{Validation}

An ISS for the ARMv5 architecture was already available in SimSoC. Thanks to
backward compatibility, all the tests running on ARMv5 can be used to test our
new ARMv6 ISS.

The new ISS passes all the tests written to validate the previous ARMv5 ISS of
SimSoC. In particular, the new ARMv6 ISS can simulate Linux running on two
boards based on the ARMv5 architecture (the SPEArPlus600 from STMicroelectronics
and the TI AM1707 from Texas Instrument).
% add result about generated decoded tests

Using a generator avoids many typo-like errors. However, other kinds of errors
remain possible. Here are the last bugs we found and fixed while trying to boot
Linux on the SPEArPlus600 SoC simulator:
\begin{itemize}
\item After the execution of an {\stt LDRBT instruction}, the content of the
  base register ({\stt Rn}) was wrong. It was due to a bug in the reference
  manual itself; the last line of the pseudo-code has to be
  deleted\footnote{This error is fixed in the ARMv7 reference manual, which is
    now the recommended manual for the ARMv6 architecture.}.
\item After a data abort exception, the base register write-back was not
  canceled, because we did not notice this rule during our first reading of the
  manual. We fix this issue as explained in section~\ref{s:writeback}.
% \item Additionally, there is an half-word access to an odd address while
%   executing SPEArPlus600 specific code. In this case, the manual indicates that
%   the result is ``unpredictable''. % ...
\end{itemize}

Once Linux was booting on the SPEArPlus600, Linux booted at the first try on the
TI AM1707 SoC simulator (the other components were already validated using the previous
hand-written ARMv5 ISS).

\subsection{Simulation Speed}

The ISS has two levels of dynamic translation. First, the instructions are
decoded and stored in an array of instruction objects. Filling this array is
quick (between 5 and 10 Mips), and then simulating one basic block is done by
calling the instructions functions one by one.
% This is the only mode available
% for the previous ARMv5 ISS.

% The new ISS is able to translate a basic block to LLVM~\cite{llvm}, and then to
% compile it to native code using the LLVM runtime compiler. This compilation is
% done only for blocks which are executed very often (>300,000 times) because this
% compilation step is very slow (<1000 Mips).

% compare with ARMv5 (backward compatibility)
% use 10 elf files and 3 computers.
We compared the speed of the generated ARMv6 ISS with the hand-written ARMv5
ISS. We wanted to know whether our approach based on extraction, transformation
and generation allows to reach the same speed ISS written and optimized by
hand.
% Thus, we made the measurement with the compilation to native code
% disabled, because it is not available in the old ARMv5 ISS.
We used three benchmarks ``loop'', ``sorting'', and ``crypto''. We compiled them
targeting either the ARM or the Thumb variant of the ARMv5 instruction set, a
first time with optimization ({\stt -O3}) and a second without ({\stt
  -O0}). Three different computers were used: a 32-bit Linux, a 64-bit Linux,
and a MacBook pro (64-bit).

\begin{table}\centering
\caption{Comparison of the simulation speeds}
\begin{tabular}{l|l|rr|r}
% \cline{3-4}
\multicolumn{2}{l|}{~} & \multicolumn{2}{|c|}{ARMv6} & \multicolumn{1}{c}{ARMv5} \\
\multicolumn{2}{l|}{~} & \multicolumn{2}{|c|}{generated ISS} & \multicolumn{1}{c}{hand-written} \\
\multicolumn{2}{l|}{~} & \multicolumn{2}{|c|}{speed and relative gain} & \multicolumn{1}{c}{speed} \\\hline
\multirow{3}{*}{arm32-crypto-O0}  & Linux 64  & 104.78 Mi/s\hspace*{-2.5mm} & +2.6\% & 102.16 Mi/s \\
                                  & MacOSX    & 89.08 Mi/s\hspace*{-2.5mm}  & +7.4\% & 82.98 Mi/s  \\
                                  & Linux 32  & 76.74 Mi/s\hspace*{-2.5mm}  &\hspace*{-2.5mm}-10.8\% & 86.03 Mi/s  \\\hline

\multirow{3}{*}{arm32-crypto-O3}  & Linux 64  & 89.97 Mi/s\hspace*{-2.5mm}  & +2.4\% & 87.89 Mi/s  \\
                                  & MacOSX    & 74.65 Mi/s\hspace*{-2.5mm}  & +4.6\% & 71.39 Mi/s  \\
                                  & Linux 32  & 70.91 Mi/s\hspace*{-2.5mm}  & -5.1\% & 74.70 Mi/s  \\\hline

\multirow{3}{*}{arm32-loop}       & Linux 64  & 124.85 Mi/s\hspace*{-2.5mm} & -1.2\% & 126.38 Mi/s \\
                                  & MacOSX    & 108.50 Mi/s\hspace*{-2.5mm} & +1.9\% & 106.52 Mi/s \\
                                  & Linux 32  & 88.89 Mi/s\hspace*{-2.5mm}  & -5.8\% & 94.39 Mi/s  \\\hline

\multirow{3}{*}{arm32-sorting-O0} & Linux 64  & 82.18 Mi/s\hspace*{-2.5mm}  & -0.5\% & 82.61 Mi/s  \\
                                  & MacOSX    & 74.40 Mi/s\hspace*{-2.5mm}  & +8.6\% & 68.49 Mi/s  \\
                                  & Linux 32  & 62.42 Mi/s\hspace*{-2.5mm}  &\hspace*{-2.5mm}-11.3\% & 70.37 Mi/s  \\\hline

\multirow{3}{*}{arm32-sorting-O3} & Linux 64  & 106.41 Mi/s\hspace*{-2.5mm} & -1.0\% & 107.54 Mi/s \\
                                  & MacOSX    & 97.51 Mi/s\hspace*{-2.5mm}  & +5.6\% & 92.35 Mi/s  \\
                                  & Linux 32  & 83.39 Mi/s\hspace*{-2.5mm}  & -1.0\% & 84.27 Mi/s  \\\hline

\multirow{3}{*}{thumb-crypto-O0}  & Linux 64  & 117.80 Mi/s\hspace*{-2.5mm} & +2.3\% & 115.15 Mi/s \\
                                  & MacOSX    & 100.22 Mi/s\hspace*{-2.5mm} & -0.5\% & 100.71 Mi/s \\
                                  & Linux 32  & 84.56 Mi/s\hspace*{-2.5mm}  & -8.1\% & 91.98 Mi/s  \\\hline

\multirow{3}{*}{thumb-crypto-O3}  & Linux 64  & 111.67 Mi/s\hspace*{-2.5mm} & +7.1\% & 104.30 Mi/s \\
                                  & MacOSX    & 98.48 Mi/s\hspace*{-2.5mm}  & +6.5\% & 92.46 Mi/s  \\
                                  & Linux 32  & 84.33 Mi/s\hspace*{-2.5mm}  & -2.9\% & 86.87 Mi/s  \\\hline

\multirow{3}{*}{thumb-loop}       & Linux 64  & 133.95 Mi/s\hspace*{-2.5mm} & +4.3\% & 128.44 Mi/s \\
                                  & MacOSX    & 108.24 Mi/s\hspace*{-2.5mm} & +3.2\% & 104.86 Mi/s \\
                                  & Linux 32  & 75.96 Mi/s\hspace*{-2.5mm}  &\hspace*{-2.5mm}-24.2\% & 100.16 Mi/s \\\hline

\multirow{3}{*}{thumb-sorting-O0} & Linux 64  & 79.61 Mi/s\hspace*{-2.5mm}  &  0.0\% & 79.61 Mi/s  \\
                                  & MacOSX    & 74.17 Mi/s\hspace*{-2.5mm}  & +1.4\% & 73.13 Mi/s  \\
                                  & Linux 32  & 62.24 Mi/s\hspace*{-2.5mm}  & -9.5\% & 68.78 Mi/s  \\\hline

\multirow{3}{*}{thumb-sorting-O3} & Linux 64  & 121.39 Mi/s\hspace*{-2.5mm} &\hspace*{-2.5mm}+26.5\% & 95.98 Mi/s  \\
                                  & MacOSX    & 97.19 Mi/s\hspace*{-2.5mm}  & +8.2\% & 89.83 Mi/s  \\
                                  & Linux 32  & 89.55 Mi/s\hspace*{-2.5mm}  &\hspace*{-2.5mm}+15.1\% & 77.81 Mi/s  \\\hline

\multirow{3}{*}{\em average}      & Linux 64  & 107.26 Mi/s\hspace*{-2.5mm} & +4.1\% & 103.00 Mi/s \\
                                  & MacOSX    &  92.24 Mi/s\hspace*{-2.5mm} & +4.5\% &  88.27 Mi/s \\
                                  & Linux 32  &  77.90 Mi/s\hspace*{-2.5mm} & -6.8\% &  83.54 Mi/s \\\hline

\multicolumn{2}{l|}{\em global average}       &  92.47 Mi/s\hspace*{-2.5mm} & +0.9\% &  91.60 Mi/s \\
\end{tabular}
\label{t:speed}
\end{table}

The results are detailed in Table~\ref{t:speed}. Globally, we obtained a small
improvement of less than 1\%. That is smaller than the measurement accuracy, and
so we can only conclude that both ISSes run roughly at the same speed. However,
we can note that the new ISS behaves better on 64-bits machine;
indeed, figures about 32-bits machines are not a real issue
because such machines become less and less common among people doing
simulation.

% add one word about DT3 speed (DT3 is not the subject of this paper)

%%%%%%%%%%%%%%%%%%%%%%%%%%%%%%%%%%%%%%%%%%%%%%%%%%%%%%%%%%%%%%%%%%%%%%%%%%%%%
\section{Conclusion}
\label{s:concl}

% reduce developpment time
% ease refactoring a lot
We have combined two techniques to generate an ISS: 1)
automatic extraction of pseudo-formal descriptions, 2) automatic analysis and
transformation of an intermediate representation of the target program. We have
obtained an ISS for ARMv6 that is as good as the previous hand-written one for ARMv5,
and the development time has been significantly reduced.
Moreover, trying a new optimization or targetting another ISS architecture
is clearly much easier with this approach.
%this approach allows to refactorize the code much more easily.

% ARMv7 doc more formal than "ARMv6" doc: fix patch problems
The effort to write our tool chain would have been even smaller if we had used
the ARMv7 reference manual. Indeed, the small bugs we noticed in the ARMv6
manual are fixed in the ARMv7 manual, and the description of the instruction set
is much more formal in the new reference manual. Thus, much of the remedial
transformation steps could be avoided.

% further work: more generic
Among the transformations steps, some are specific to the ARM architecture, but
others could be reused for other architectures such as MIPS or PowerPC. Reusing
code would require to agree on an abstract architecture-independent language,
and to group the interesting functions in a library.

% back-ends
We have currently three back-ends: the fast ISS back-end, the tests for the
decoder, and the Coq formal specification. Moreover, we have almost finished
another backend allowing LLVM-based dynamic translation.
% We intend to prove that some key
% parts of the fast ISS are consistent with the Coq specification --
% that would require to generate part of Coq proofs.
%Moreover,
The intermediate representation contains data that could be useful to generate
an assembler.  We could also generate descriptions in other ADL languages, and
use the associated tools.

%%%%%%%%%%%%%%%%%%%%%%%%%%%%%%%%%%%%%%%%%%%%%%%%%%%%%%%%%%%%%%%%%%
\bibliographystyle{IEEEtran}
\bibliography{simlight2}

\end{document}